\documentclass[nofootinbib,twocolumn,aps,pre,superscriptaddress,citeautoscript,floatfix]{revtex4-1}
\setcitestyle{super}
\usepackage{graphics}
\usepackage{graphicx}
\usepackage{mathrsfs}
\usepackage{amsmath}
\usepackage{bm}
\usepackage{color}
\usepackage{float}

\begin{document}

\title{Effect of Sinusoidal Surface Roughness and Energy on the Orientation of Cylinder-Forming Block Copolymer Thin Films}
\author{Yanyan Zhu}
\affiliation{Center of Soft Matter Physics and its Applications, Beihang University, Beijing 100191, China}
\affiliation{School of Physics and Nuclear Energy Engineering, Beihang University, Beijing 100191, China}
\author{Karim Aissou}
\affiliation{Institut Europ\'een des Membranes, Universit\'e de Montpellier-CNRS-ENSCM, 300 Avenue du Professeur Emile Jeanbrau, F-34090 Montpellier, France}
\author{David Andelman}
\affiliation{Raymond and Beverly Sackler School of Physics and Astronomy, Tel Aviv University, Ramat Aviv 69978, Tel Aviv, Israel}
\author{Xingkun Man}
\email{manxk@buaa.edu.cn}
\affiliation{Center of Soft Matter Physics and its Applications, Beihang University, Beijing 100191, China}
\affiliation{School of Physics and Nuclear Energy Engineering, Beihang University, Beijing 100191, China}

\begin{abstract}

We explore the relative stability of three possible orientations of cylinder-forming di-block copolymer on a sinusoidally corrugated substrate. The cylinders can be aligned either parallel to the substrate, with their long axis being oriented along or orthogonal to the corrugation trenches, or perpendicular to the substrate. Using self-consistent field theory, we investigate the influence of substrate roughness and surface preference on the phase transition between the three orientations. When the substrate preference, $u$, towards one of components is small, increasing the substrate roughness induces a phase transition from parallel to perpendicular cylindrical phase. However, when $u$ is large, the parallel orientation is more stable than the perpendicular one. Within this parallel phase, increasing the substrate roughness leads to a transition of cylinder orientation changing from being orthogonal to parallel to the trench long axis. Increasing the substrate preference leads to an opposite transition from parallel to orthogonal to the trenches. Furthermore, we predict that the perpendicular cylinder phase is easier to be obtained when the unidirectional corrugation is along the longer unit vector of the hexagonal packing than when it is along the shorter unit vector. Our results qualitatively agree with previous experiments, and contribute towards applications of the cylinder-forming block copolymer in nanotechnology.

\end{abstract}

\maketitle

\section{Introduction}

Block copolymers (BCPs) are composed of two or more chemically distinct blocks, which are covalently bonded together. The chemical incompatibility between the different blocks drives a microphase separation, in which the BCP can form a variety of well-ordered nanostructures via self-assembly. The phase behavior of BCP melts has been studied extensively in recent years, showing a rich variety of morphologies, such as lamellae, hexagonally close-packed (HCP) cylinders, body-centered cubic (BCC) packing of spheres, and complex networks such as the cubic double gyroid (Q$^{230}$) and orthorhombic O$^{70}$ phases~\cite{Bates94,Epps04,Li13,Xie14}. The length scale of microphase separation is in the range of 10-100 nm, making them ideal for emerging nanotechnologies~\cite{Bates99}, including applications in nanolithography~\cite{cheng01,Jeong08}, nanoporous membranes~\cite{Yang06} and magnetic nanowires~\cite{Thurn00,Peng11}.

In many cases, capturing the vast technological potential of BCP thin film requires precise control over the orientation and the lateral alignment of these nanostructures in order to produce defect-free array of BCP features. In recent decades, much effort has been devoted to tailor the self-assembly behavior of BCP thin films by using engineering surface effects~\cite{Huang98}, external fields~\cite{Morkved96}, patterned substrates~\cite{Kang12,Vega13a,Zhang14,Cong16}, and solvent vapor annealing~\cite{Sinturel13}. Among these approaches, the use of nonflat substrates to direct the self-assembly of BCP thin films has been proven to be an effective method to achieve long-range ordered arrays with either a parallel or perpendicular orientation of BCP domains with respect to the substrate.

When BCP lamellae or cylinders are parallel to the unidirectional corrugated substrate, the domain orientation can be orthogonal, parallel or aligned with a tilted angle with respect to the trench long axis. Previous experimental studies have shown that the film thickness~\cite{Hong12,Hong2012,Choi18a}, substrate mean curvature~\cite{Vega13a} and roughness and chemical preference of the substrate~\cite{Tavakkoli16} are key factors in determining the orientation of BCP domains with respect to the trenches.

A large lateral scale and nearly defect-free cylindrical BCP thin films, which are perpendicular to the substrate, were obtained for various types of nonflat substrates, such as sawtoothed topography~\cite{Park09}, sinusoidal pattern~\cite{Aissou13,Rho15}, ordered nanoparticle monolayers~\cite{Kim13a}, minimal topographic pattern~\cite{Choi16} etc. To the best of our knowledge, only Kim \emph{et al.}~\cite{Kim13a} and Aissou \emph{et al.}~\cite{Aissou15} investigated experimentally the transition of cylinder orientation from parallel to perpendicular with respect to the nonflat substrate. They showed that this transition can be obtained either by increasing the substrate roughness~\cite{Kim13a} or decreasing the film thickness~\cite{Aissou15}.

In view of the above mentioned experimental studies, there are only a few theoretical works addressing the self-assembly of BCP films on corrugated surfaces. Peng \emph{et al.}~\cite{Peng15} employed self-consistent field theory (SCFT) to explore the self-assemble behavior of cylinder-forming BCP thin films on a saw-toothed substrate. They investigated the effects of the substrate corrugation periodicity and the film thickness on cylindrical structures. Man \emph{et al.}~\cite{Man15,Man16} systematically studied the self-assembly of lamellar forming BCP thin film on a sinusoidal substrate, and showed an enhanced synergy between substrate topography combined with a weak surface preference to obtain defect-free perpendicular lamellar BCP thin films. Recently, Carpenter \emph{et al.}~\cite{Carpenter17} presented a study combining SCFT calculations with experimental results, and found that the orientation of cylinders with respect to the trench depends on the commensurability of the BCP hexagonal packing with the substrate characteristic length and film thickness.

Aforementioned studies~\cite{Peng15} show that when cylinder forming BCP self-assembly on an unidirectional corrugated substrate, there are three possible orthogonal cylinder orientations. The cylinders can be either perpendicular or parallel to the substrate, while the latter one can be characterized by a titled angle with respect to the trench long axis. Previous studies~\cite{Aissou15,Choi18a} showed that varying the substrate roughness and the film thickness can cause phase transitions between these orientations. In spite of this progress, a quantitative mechanistic understanding of the effect of non-flat substrate in determining the cylinder orientation in BCP thin films is still missing. Here, we investigate the self-assembly of cylinder-forming di-BCP thin films on sinusoidally corrugated substrate. By using SCFT, our aim is to explain the effects of substrate geometry and relative surface preference to one of the di-BCP components on the transition between the three cylinder orientations.

This paper is organized as follows. In the next section, we introduce the self-consistent field theory (SCFT) technique and our model. In Sec. III, we present the phase diagram of BCP cylinder on corrugated substrates, followed by the discussion of our results in Sec. V. Finally, Sec. IV, we present our conclusions and some future prospects.

\begin{figure*}
{\includegraphics[width=0.6\textwidth,draft=false]{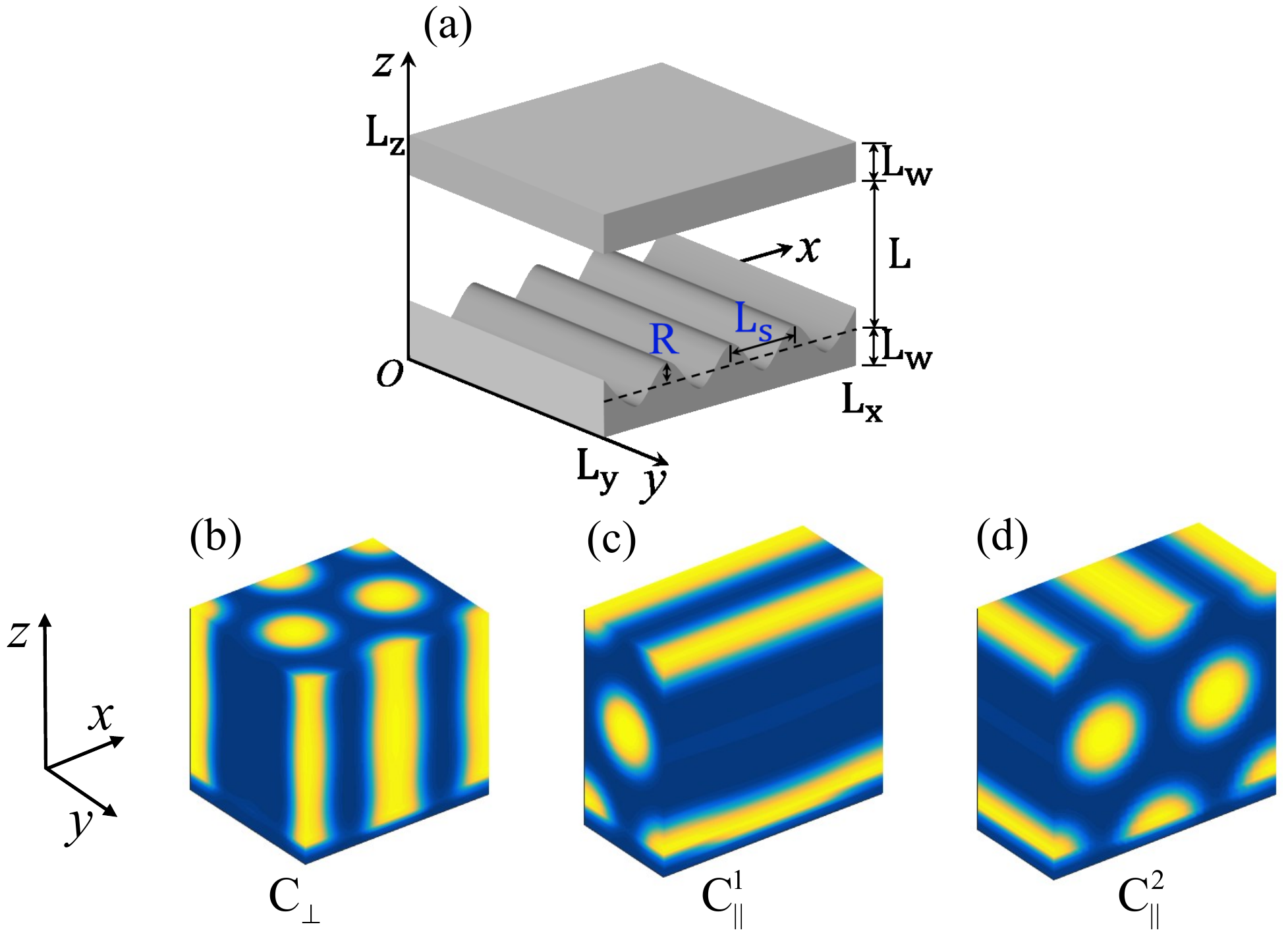}}
\caption{
\textsf{Schematic illustration of a BCP film confined between two surfaces and its cylinder orientations for a bottom substrate that is sinusoidally corrugated. The size of the 3D calculation box
is $L_x\times L_y \times L_z$. (a) The averaged BCP film thickness is $L=L_z-2L_w$, where $L_w$ is the average wall thickness within the box. The substrate corrugation describes trenches that are translationally invariant in the $y$-direction with periodicity $L_s$ and amplitude $R$, $h(x)=R\cos(2\pi x/L_s)$. A-rich regions are denoted by yellow and B-rich by blue. In (b), the cylinder orientation is perpendicular to the substrate, and the phase is denoted as C$_{\perp}$, while in (c) and (d) we show the two orientations that are parallel to the substrate (and to each other). In (c), the cylinder  are perpendicular to the long axis of the substrate trenches and the phase is denoted as C$^{1}_{\parallel}$. In (d), the cylinder are  parallel to the trench long-axis and the phase is denoted as C$^{2}_{\parallel}$.}}
\label{fig1}
\end{figure*}

\section{Model}

We employ self-consistent field theory (SCFT) to investigate the self-assembly of cylinder-forming block copolymer thin film. The BCP thin film is confined between a flat top surface and a sinusoidally corrugated bottom one, as showed in Figure~1(a). Specifically, we describe the polymer as a Gaussian chain composed of $N$ segments, of which a fraction $f$ are of type A and $(1{-}f)$ of type B. The interactions between A and B monomers are mediated through the Flory parameter $\chi_{\rm AB}$, and $u=N\chi_{{\rm sA}}-N\chi_{{\rm sB}}$ is the relative interaction between the substrate and the A (B) component, where $\chi_{{\rm sA}}$ ($\chi_{{\rm sB}}$) is the interaction parameter between the substrate and the A (B) component. This choice means that $u>0$ induces substrate preference of the A component. We model the periodic surface trenches by a single $q$-mode along the $x$-direction with periodicity $L_s$ and amplitude $R$, $h(x)=R\cos(2\pi x/L_s)$. Lateral confinement is modeled using the masking method, where the wall is described as the third component~\cite{Matsen97}. All lengths are rescaled with the chain radius of gyration, $R_g=\sqrt{Nb^2/6}$, where $b$ is the Kuhn length taken for simplicity to be the same for the two blocks.

The Hamiltonian for a di-BCP film confined between the two surfaces can be expressed as a functional of two conjugate potential fields, $W_+(x)$ and $W_-(x)$
\begin{equation}
\begin{aligned}
&H[W_+,W_-]=C\int d^3\bm{r} \left( \frac{[W_-(\bm{r})]^2}{N\chi_{{\rm AB}}}-\frac{2Nu}{N\chi_{{\rm AB}}}\phi_w(\bm{r})W_-(\bm{r})
\right.\\
&\left. +\frac{[W_+(\bm{r})]^2-2\zeta N\phi_p(\bm{r})iW_+(\bm{r})}{N\chi_{{\rm AB}}+2N\zeta} \right) -C\Omega\bar{\phi_p}\ln Q[W_{\rm A},W_{\rm B}]
\end{aligned}
\end{equation}
where $C=\rho_0 R_g^3/N$ is a normalization factor. The total volume of the simulation box is $\Omega$, and $\phi_w+\phi_p=1$, $\phi_w$ is the wall volume fractions and $\phi_p(\bm{r})$ is the dimensionless volume fraction of the polymer, $\phi_p(\bm{r})=\phi_A(\bm{r})+\phi_B(\bm{r})$. $\bar{\phi_p}=\Omega ^{-1}\int d^3\bm{r}\;\phi_p(\bm{r})$ is the polymer volume fraction averaged over $\Omega$. In addition, $\zeta$ is a penalty cost for local density deviation from the incompressibility condition, and $Q[W_A,W_B]=\Omega^{-1}\int d^3\bm{r}q(\bm{r},s{=}1)$ is the single-chain partition function for BCP, in which the propagator $q(\bm{r},s)$ is the solution of the following modified diffusion equation
\begin{equation}
\frac{\partial q(\bm{r},s)}{\partial s}=\nabla ^2 q(\bm{r},s)-W(\bm{r},s)q(\bm{r},s)
\end{equation}
where $W(\bm{r})=W_A(\bm{r})$ for $0\le s<f$ and $W(\bm{r})=W_B(\bm{r})$ for $f\le s\le 1$. The initial condition for eq 2 is $q(\bm{r},s)=1$.

In the mean-field approximation, the thermodynamic properties of the confined melt can be obtained from saddle-point configurations of the Hamiltonian in eq 1, {\it i.e.}, solutions of
\begin{equation}
\frac{\delta H[W_+,W_-]}{\delta (iW_+(\bm{r}))}=\frac{\delta H[W_+,W_-]}{\delta (W_-(\bm{r}))}=0
\end{equation}
A detailed formulation of the numerical procedure and its implementation to SCFT modeling of BCP systems can be found elsewhere~\cite{Bosse07,Hur09,Takahashi12}.

The SCFT formulation gives the local density for the A and B components, $\phi_{\rm A}(\mathbf{r})$ and $\phi_{\rm B}(\mathbf{r})$, respectively. There are three orientations of the cylindrical phase with respect to the substrate, as shown schematically in Fig.~1. The perpendicular orientation is denoted C$_{\perp}$ (Fig.~1(b)), while the parallel orientation can be divided into two orientations, C$^{1}_{\parallel}$ and C$^{2}_{\parallel}$, which are orthogonal to each other, as well as to the C$_{\perp}$, as shown in Fig.~1(c) and (d). Whereas C$^{1}_{\parallel}$ is orthogonal to the trench long-axis, the C$^{2}_{\parallel}$ ordering is oriented along the trench direction. The BCP film is in contact with a uni-axial corrugated substrate, simulated which has a small preference toward one of the two BCP components.

\bigskip\bigskip

\begin{figure}
{\includegraphics[width=0.4\textwidth,draft=false]{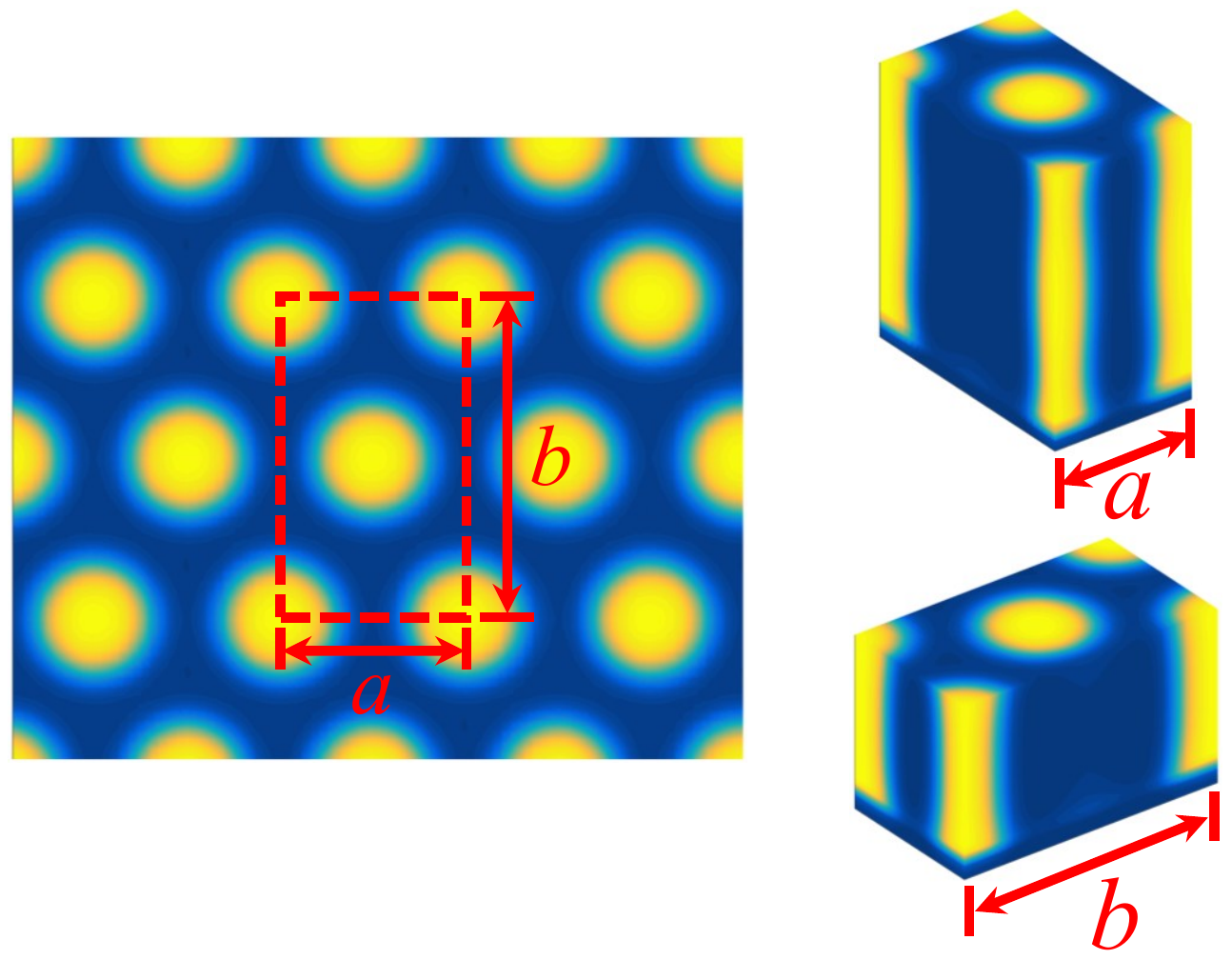}}
\caption{
\textsf{Schematic drawing of the top-view of the hexagonally packed cylindrical phase (left). The two directions, labeled as $a$ and $b$, indicate the short and long unit vectors of the hexagonal unit cell, respectively. The unidirectional corrugated substrate undulates along the $a$-direction or the $b$-direction of the cylindrical phase, as shown in the right of the figure.
 }}
\label{fig2}
\end{figure}

From the characteristics of the hexagonal phases as shown in Fig.~2, it is evident that a unidirectional corrugated substrate undulate not only along the short ($a$-direction) but also along the long ($b$-direction) unit vectors. Therefore, we present hereafter the effect of the substrate on the cylindrical phase in both cases, and the corresponding phase diagram of the three orientations.

\section{Results}

\begin{figure*}
{\includegraphics[width=0.65\textwidth,draft=false]{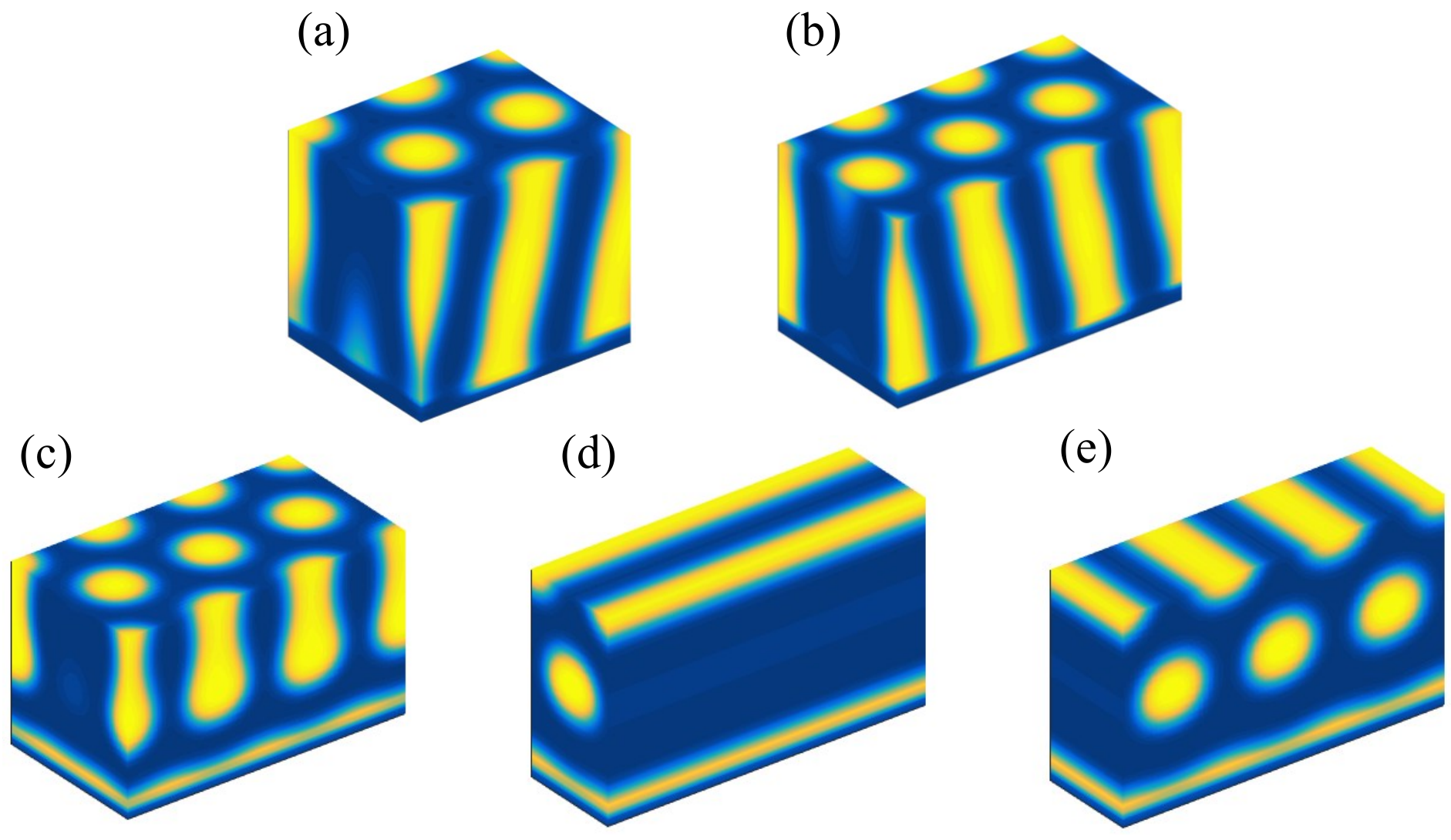}}
\caption{
\textsf{SCFT calculation of BCP cylindrical phase in contact with a substrate characterized by $a$, small $L_s$ in (a), large $R$ in (b) and large $u$ in (c-e). For (a), the corrugation periodicity is $L_s=2L_0$ and its amplitude $R=0.3$, leading to a deformed perpendicular cylindrical phase, C$_{\perp}$. For (b), the substrate parameters are $L_s=3L_0$ and $R=0.4$ also resulting in a deformed C$_{\perp}$. In both (a) and (b), the top and bottom surfaces are neutral. Strong substrate preference results in deformed cylindrical phases for all three orientations (C$_{\perp}$, C$^1_{\parallel}$, C$^2_{\parallel}$), as shown in (c), (d), and (e), respectively, and other parameters are $R=0$ and $u=9$.
 }}
\label{fig3}
\end{figure*}

We focus on the effects of sinusoidal substrates on the orientation of a confined BCP cylindrical phase. The Flory-Huggins interaction parameter is taken as $N\chi_{\rm AB}=25$, and the fraction of minority A-component is $f=0.3$. For these values, the behavior falls well within the cylindrical region of the bulk phase diagram~\cite{Matsen12}. For this $N\chi_{\rm AB}$ value, the characteristic BCP lengths in a thin film geometry are $L^{a}_0=4.4$ and $L^{b}_0=\sqrt{3}L^{a}_0=7.6$ (in units of $R_g$). These values are obtained by varying the film thickness and comparing the corresponding free energies, in order to find at which thickness the free energy has a minimum. We set the average film thickness to be an integer number of $L^{b}_0$ ($L^{a}_0$) in order to limit the $z$-direction space confinement effects on the cylinder orientation when the unidirectional corrugation is along the $a$-direction ($b$-direction). The top surface is always flat and neutral ($u_{\rm top}=0$). The orientation of the cylindrical phase in the BCP thin film mainly depends on the substrate roughness whose height, $h(x)=R\cos(2\pi x/L_s)$, is described by the corrugation periodicity $L_s$ and amplitude $R$. The strength of substrate preference towards one of the two components is $u$, and is chosen to be a positive when the substrate prefers the minority A-component.

Figure~3 shows various deformed BCP cylindrical phases due to either large substrate roughness or strong  substrate preference. For deformed cylindrical phases, it is hard to recognize which orientation is the equilibrium structure. Therefore, we limit ourselves to a range of parameters resulting in perfect BCP cylindrical phase. Figure~3a is a deformed C$_{\perp}$ phase on a neutral substrate with $L_s=2L^{a}_0$ and $R=0.3$. We find that when $L_s\le 2L^{a}_0$, it is difficult to obtain perfect cylindrical phase even when $R$ is reasonably small. Furthermore, the value of $R$ cannot be too large, otherwise C$_{\perp}$ is deformed as shown in Figure~3b. Numerical calculations show that $R<0.4$ for $L_s=3L^{a}_0$ and for $u=0$, in order to avoid such deformations. Besides the substrate roughness, the strength of the substrate preference, $u$, can also induce deformed cylindrical phases. Figures~3c-e show that $u=9$ is already large enough to generate a wetting layer of A-component along the substrate surface. Therefore, all our simulations were conducted within the parameter range of $R<0.4$, $L_s\ge3L_0$ and $|u|\le9$.

\begin{figure*}
{\includegraphics[width=0.7\textwidth,draft=false]{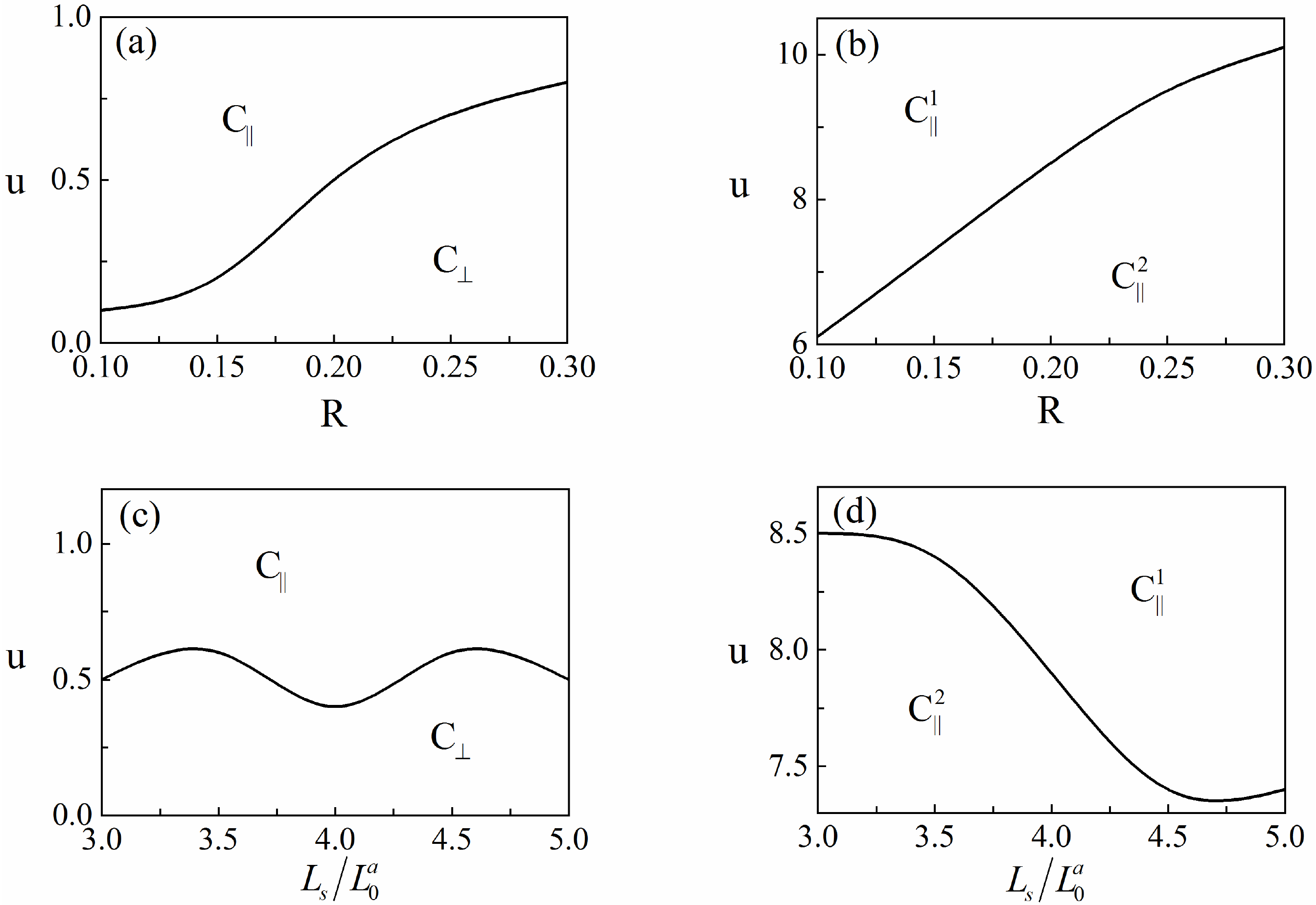}}
\caption{
\textsf{C$_{\perp}$-to-C$_{\parallel}$ and C$^2_{\parallel}$-to-C$^1_{\parallel}$ phase transitions in terms of the substrate roughness parameter: the amplitude $R$, the ratio $L_s/L_0$, and preference $u$. The lines separate the parallel orientations above from the perpendicular orientation below. (a) The C$_{\perp}$-C$_{\parallel}$ phase-transition in the ($R$, $u$) plane. (b) The phase transition of the two parallel orientations, separating the C$^1_{\parallel}$ above from C$^2_{\parallel}$ below. In both (a) and (b), $L_s=3L^a_0$. (c) and (d) show the same phase diagrams as in (a) and (b), separately, but for the ($L_s/L_0$, $u$) plane, where the substrate roughness is varied by changing $L_s$, while keeping $R=0.2$.  For all cases, the top flat surface is neutral, $N\chi_{\rm AB}=25$ and $L^a_0=4.4R_g$.
}}
\label{fig4}
\end{figure*}

\subsection{The unidirectional substrate corrugation along the $a$-direction}

The unit cell of the cylindrical phase is composed of two unit vectors: a short one, $L^a_0$, and the long one, $L^b_0$. The effects of non-flat substrate on the relative stability of the cylindrical phase with three orthogonal orientations (C$_{\perp}$, C$^{1}_{\parallel}$, C$^{2}_{\parallel}$) will be different when the substrate unidirectional corrugation undulates along the $a$ or $b$ direction. This is due to the fact that the distortion effects of nonflat substrates on the BCP cylinders mainly depends on the ratio between the periodicity of sinusoidal substrate and the cylinder characteristic lengths.

We start with the corrugation along the $a$-direction, and study quantitatively the effect of corrugated substrate on the phase transition between the three orientations. We focus on the role played by (i) the substrate roughness that is described by lateral variational periodicity $L_s$ and roughness amplitude $R$; and, (ii) the relative surface preference toward BCP components, $u$.

Figure~4 shows the phase transition between perpendicular and parallel orientations (C$_{\perp}$-to-C$_{\parallel}$), and between the two parallel orientations themselves (C$^{2}_{\parallel}$-to-C$^{1}_{\parallel}$) in terms of the substrate roughness, $2\pi R/L_s$, and the substrate preference, $u$. In previous studies~\cite{Man15}, it has been shown that there are two ways to change the substrate roughness. First we fix the $L_s=3L^{a}_0$ and then change the amplitude $R$ from $0.1$ to $0.3$, as shown in Figures~4a and 4b. Alternatively, we present in Figures~4c and 4d the cases where $L_s/L^{a}_0$ varies from $3$ to $5$ while keeping $R=0.2$. For both cases, we investigate weak substrate preference (small $u$) and strong substrate preference (large $u$), separately. The top surface is always taken to be a neutral surface.

The substrate preference makes the parallel orientation more stable than the perpendicular orientation. Therefore, increasing $u$ results in a C$_{\perp}$-to-C$_{\parallel}$ phase transition as shown in Figures~4a and 4c. Here, C$_{\parallel}$ includes both C$^{1}_{\parallel}$ and C$^{2}_{\parallel}$. It is interesting to note that when $u$ becomes large (Figures~4b and 4d), the transition of C$^{2}_{\parallel}$-to-C$^{1}_{\parallel}$ is obtained, indicating that C$^{1}_{\parallel}$ is more stable than C$^{2}_{\parallel}$ when the substrate preference is strong. The critical $u$ value that is needed to induce both the C$_{\perp}$-to-C$_{\parallel}$ and C$^{2}_{\parallel}$-to-C$^{1}_{\parallel}$ transitions increases function of substrate roughness (see Figure 4a, b, and d). However, the critical $u$ value oscillates when $R$ is fixed and $L_s/L^{a}_0$ = $3$, $3.5$, $4$, and $5$ as shown in Figure~4c. Figure~4c also shows that for weak preferences, the critical $u$ value is smaller when $L_s/L^{a}_0$ is an integer number, as compared with half integer values of $L_s/L^{a}_0$. It is clear that the distortion is smaller in the former case. For strong $u$ preferences, the substrate preference dominates the relative stability of the two parallel orientations. Therefore, $u$ is a monotonically increasing function of the substrate roughness, as shown in Figure~4b and 4d.

\begin{figure*}
{\includegraphics[width=0.9\textwidth,draft=false]{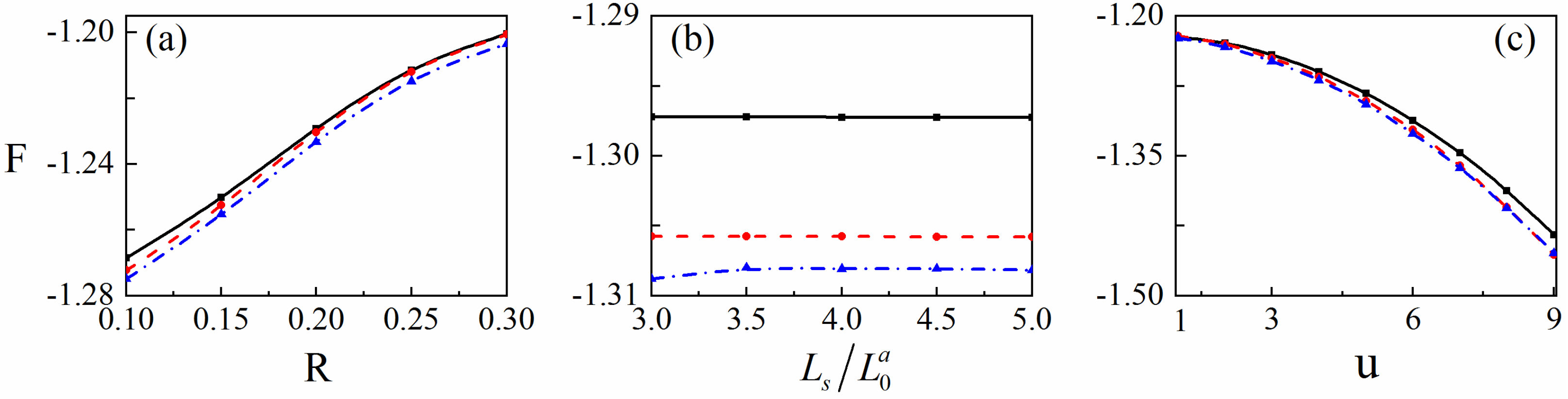}}
\caption{
\textsf{The dependence of the free energy for the three orientations C$_{\perp}$ (black line), C$^{1}_{\parallel}$ (red line), C$^{2}_{\parallel}$ (blue line) on (a) substrate corrugation amplitude, $R$, (b) rescaled substrate lateral corrugation periodicity, $L_s/L^{a}_0$, and (c) substrate preference, $u$. The top surface is neutral, $u_{\rm top}=0$, and $N\chi_{\rm AB}=25$, and the unidirectional substrate corrugation is along the $a$-direction.
}}
\label{fig5}
\end{figure*}

In order to understand the effects of non-flat substrates on the relative stability of the three orientations, C$^{1}_{\parallel}$, C$^{2}_{\parallel}$, and C$_{\perp}$, we study the dependence of their corresponding free energy on $R$, $L_s$, and $u$. In Figure~5a $u=2$ and all other parameters are the same as in Figure 4b. It is clear that the free energies for the three orientations increase as $R$ increases due to the increase of substrate roughness. Moreover, the two parallel orientations, C$^{1}_{\parallel}$ and C$^{2}_{\parallel}$, are more stable than the perpendicular orientation because of the surface preference. On the other hand, all free energies are nearly unchanged when $L_s$ increases, as shown in Figure~5b for $u=2$ and $R=0.2$. This occurs only because we scan a limited $L_s$, range $3\le L_s/L^{a}_0\le 5$ in order to avoid deformations of the cylindrical phase(as in Figure~3). It is also found that the stability range of C$^{2}_{\parallel}$ is larger than that of C$^{1}_{\parallel}$ and C$_{\perp}$, and the latter C$_{\perp}$ phase is the most unstable one. These results agree with the phase diagram shown in Figures~4b and 4d.

The dependence of the free energy on the surface $u$ preference is shown in Figure~5c, where calculations are done for $R=0.2$, $L_s/L^{a}_0=3$, and a neutral top surface. For the three cases, the free energy decreases as $u$ increases, and both C$^{1}_{\parallel}$ and C$^{2}_{\parallel}$ orientations free energies decrease faster than the perpendicular one. This is consistent with the results shown in Figure 4, and also with previous studies~\cite{Man15,Man16,Tavakkoli16}.

\begin{figure}
{\includegraphics[width=0.35\textwidth,draft=false]{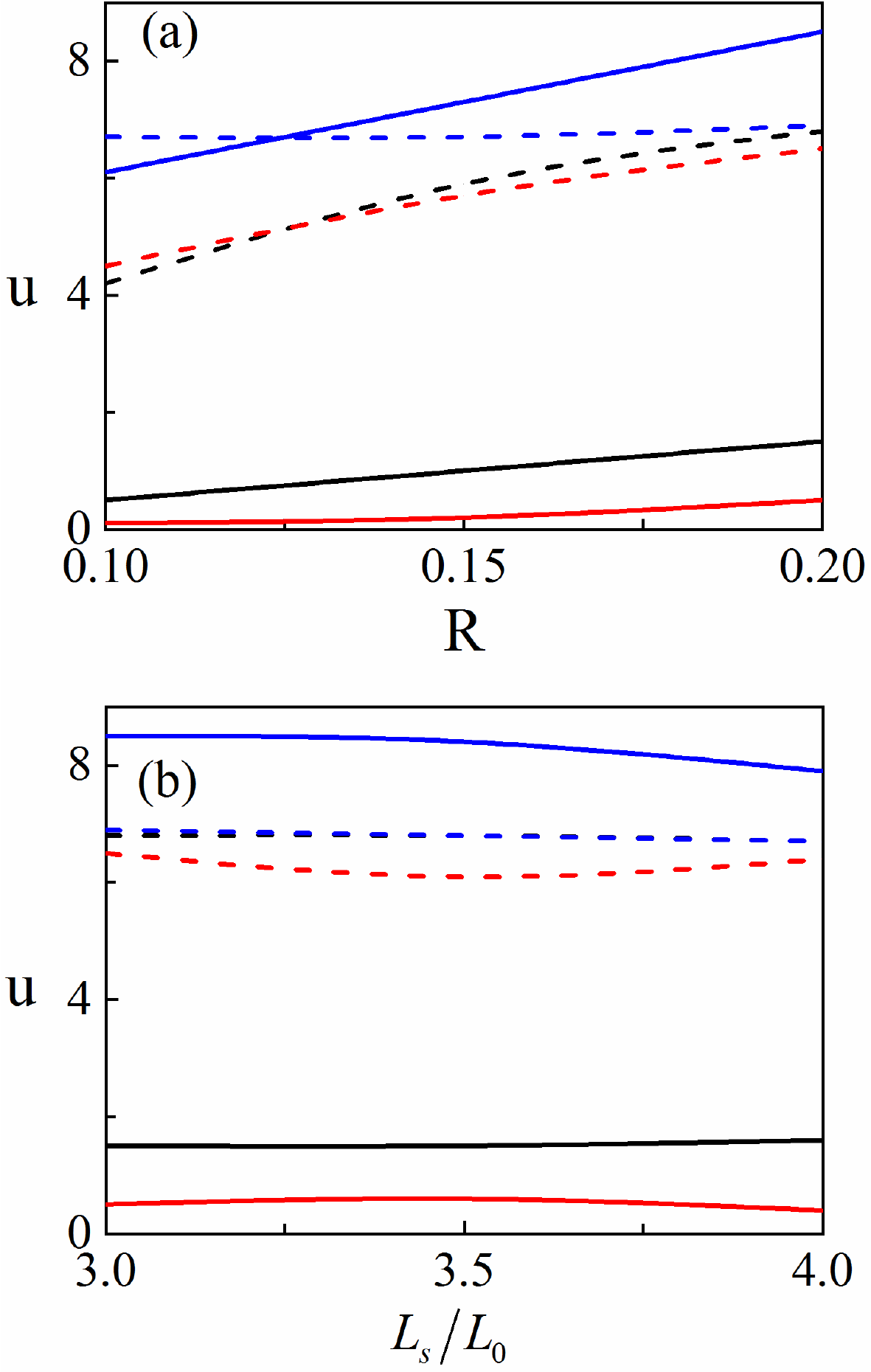}}
\caption{
\textsf{Comparison of the relative stability of the three orientations (C$_{\perp}$, C$^1_{\parallel}$, C$^2_{\parallel}$). The substrate corrugation along the $a$- and $b$- directions are represented as the solid and dashed lines, respectively. (a) Phase diagrams in the ($R$, $u$) plane for $L_s=3L^a_0$. (b) Phase diagrams in the ($L_s/L_0$, $u$) plane for $R=0.2$. The top surface is neutral for both cases.
}}
\label{fig6}
\end{figure}

\subsection{The unidirectional substrate corrugation along the $b$-direction}

Hereafter, we investigate the same phase diagram as in Figure~4, but with the substrate corrugation now along the $b$-direction. We scan a smaller range of the $R$ and $L_s$ parameters, because the characteristic length, $L^b_0=\sqrt{3}L^a_0=7.6$, is larger than $L^a_0$ in the $a$-direction. In order to have $L_s/L^b_0$ large enough to maintain a perfect cylindrical phase ($L_s/L^{b}_0 \ge 3$), the size of the calculation box in the $x$-direction should be much larger than those presented in Figure~4 for the same ratio. Consequently, we take three values of $L_s/L^{b}_0=3$, $3.5$, and $4$ for constant $R=0.2$, and for a fixed value $L_s=3L^{b}_0$, $R=0.1$, $0.15$, and $0.2$.

A comparison of the phase diagram of C$_{\perp}$-C$_{\parallel}$ and C$^{2}_{\parallel}$-C$^{1}_{\parallel}$ for the two different corrugation directions is presented in Figure~6, where the solid and dashed lines correspond to the corrugation along $a$-direction and $b$-direction, respectively. The critical value of $u$ needed to induce the C$_{\perp}$-to-C$_{\parallel}$ transition is much larger for the corrugation along $b$-direction than along the $a$-direction, regardless of the substrate roughness values (either by varying $R$ or $L_s$). This is shown in Figures~6a and 6b. Our result indicates that the perpendicular orientation is more stable when it is in contact with a unidirectional substrate corrugation along the $b$-direction rather than along the $a$-direction. We further note that the behavior of the C$^{2}_{\parallel}$-to-C$^{1}_{\parallel}$ transition in terms of the substrate roughness and preference $u$ is nearly the same for both $a$- and $b$- substrate corrugation directions. The transition between the C$^{1}_{\parallel}$ and C$^{2}_{\parallel}$ phases takes place only when $u$, the substrate preference, is strong. Therefore, the relative stability of the two parallel orientations is mainly dominated by the substrate preference, and not by the substrate roughness.

Although the substrate corrugation direction has a large effect on the relative stability of parallel and perpendicular cylinders, the corresponding free energy dependence on $R$ and $u$ is quite similar. Figure~7 shows that the free energy of the three orientations increases as a function of $R$, but decreases with $u$. Moreover, the free energy nearly remains unchanged when $L_s/L^b_0$ changes from $3$ to $4$, in analogy with the corrugations along $a$-direction.

\bigskip\bigskip

\begin{figure*}
{\includegraphics[width=0.95\textwidth,draft=false]{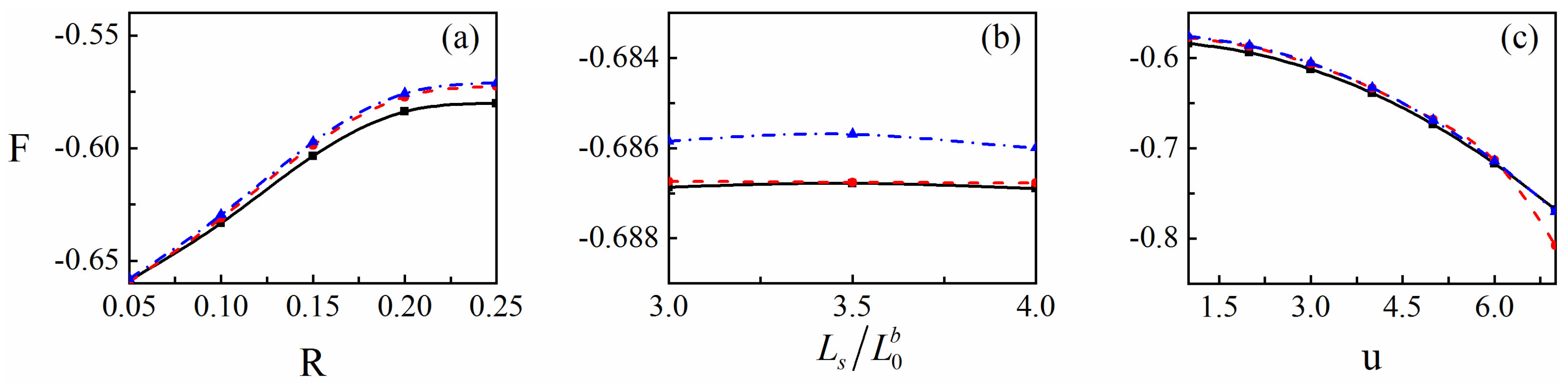}}
\caption{
\textsf{The dependence of the free energy for the three orientations: C$_{\perp}$ (black line), C$^{1}_{\parallel}$ (red line) and C$^{2}_{\parallel}$ (blue line) on (a) substrate corrugation amplitude $R$; (b) rescaled substrate lateral corrugation periodicity $L_s/L^b_0$; and, (c) substrate preference $u$, for the unidirectional substrate corrugation along the $b$-direction. The top surface is neutral and $N\chi_{\rm AB}=25$.
}}
\label{fig7}
\end{figure*}

\section{Discussion And Conclusions}

We explore how nonflat substrates affect the relative stability between the three orientations, C$_{\perp}$, C$^{1}_{\parallel}$ and C$^{2}_{\parallel}$, of BCP cylindrical phases. In general, when $u$ is small, increasing the substrate roughness will enhance the compression of the parallel cylinders because the polymers stand perpendicularly with respect to the substrate. Therefore, the perpendicular cylinder phase becomes more stable than C$_{\parallel}$ phase due to the fact that the polymers are lying down and feel less compression in C$_{\perp}$ phase. For large $u$, C$_{\parallel}$ becomes more stable than C$_{\perp}$, because the substrate has a strong preference to the A-block. In the C$_{\parallel}$ stable region of the phase diagram, increasing the substrate roughness causes a transition from C$^{1}_{\parallel}$ to C$^{2}_{\parallel}$. This happens because the corrugation causes larger distortion to the C$^{1}_{\parallel}$ phase than to the C$^{2}_{\parallel}$ one. The former one aligns orthogonally to the trenches, while the latter one aligns along with the trenches. Moreover, increasing $u$ leads a transition from C$^{2}_{\parallel}$ phase to C$^{1}_{\parallel}$ one. Here, the contact area of the A-component with the substrate in the C$^{1}_{\parallel}$ phase is the largest among the three phases.

Corrugated substrates affect differently on the relative stability between C$_{\perp}$ and C$_{\parallel}$ when the corrugation is along $a$- or $b$-direction. For substrate corrugated along the $b$-direction, we can see that $L^b_0$ is larger than $L^a_0$, resulting in a smaller substrate roughness. Hence, the corrugation perturbations to the cylinder phase are relatively small. This is why the perpendicular orientation is more stable when the substrate corrugation is along the $b$-direction rather than along the $a$-direction. However, we also note that the sinusoidal substrate-induced transition between the two parallel phases is roughly the same when the corrugation in either of the two directions.

We can obtain a stable perpendicular cylindrical phase by varying the substrate roughness parameters $R$ or $L_s$. As shown in Figure~4a, increasing $R$ causes the perpendicular orientation to become more stable for $R<0.4$, $L_s=3L^{a}_0$ and $|u|\le1$. This observation agrees well with Kim \emph{et al.} experiment~\cite{Kim13a}. They investigate the domain orientation of thin films of polystyrene-block-poly (methyl methacrylate) (PS-b-PMMA) placed on manolayer of ordered nanoparticle (NP). A transition from parallel to perpendicular orientation was obtained by increasing the substrate roughness with changing the NP diameter from 6 nm to 22 nm. In addition, Aissou \emph{et al.}~\cite{Aissou15} reported that the domain orientation can be controlled by tuning the layer thickness of Poly(1, 1-dimethyl silabutane)-b-PMMA (PDMSB-b-PMMA) deposited on a topographical varying substrate. The C$_{\parallel}$ orientation was obtained when the film thickness was 40 nm, while for film thickness of 30 nm the orientation C$_{\perp}$. This indicated that the perpendicular C$_{\perp}$ phase is more stable when decreasing the thickness of thin BCP films. For confined di-BCP thin film, the effect of decreasing the film thickness is equivalent to increasing the roughness of the substrate. This was indeed shown by Vu \emph{et al.} simulations~\cite{Vu18}, who demonstrated via SCFT calculation that the film thickness is a decreasing function of the mean curvature.

Furthermore, as can be seen in Figure~4b and 4d, for large $u$, increasing the substrate roughness results in a C$^{1}_{\parallel}$-to-C$^{2}_{\parallel}$ transition. Choi and co-workers~\cite{Choi18a} observed a similar phenomenon through changing the film thickness. They show scanning force microscopy (SFM) images of PS-b-poly (ethylene oxide) (PS-b-PEO) thin films on the minimal patterns. As the film thickness increased from 22.6 nm to 41.0 nm, the cylindrical microdomains oriented from aligning parallel to orthogonal to the trench direction (C$^{2}_{\parallel}$-to-C$^{1}_{\parallel}$). If the increase of film thickness can be thought of as reducing the substrate roughness, our findings qualitatively agree with those experimental findings.

Varying the preference of the substrate for the minority A-component can also cause a phase transition between the three cylindrical orientations. The C$^{1}_{\parallel}$ is more stable for large $u$ values. Similar effects of $u$ were reported by Man \emph{et al.}~\cite{Man15,Man16} They showed that increasing $u$ can cause a phase transition from perpendicular to parallel orientation of lamellar phases. This indeed indicates that for both cylinder and lamellae, the substrate preference $u$ can cause a phase transition from perpendicular to parallel orientation.

We illustrate, in addition, the relative stability by comparing the free energy of the three orientations. The dependence of the corresponding free energies is shown in Figure~5 and 7. Our results demonstrate that the free energy is an increasing function of the substrate amplitude $R$. This is consistent with previous simulations~\cite{Vu18}, which showed that the free energy is an increasing function of the substrate curvature. In another study, Peng \emph{et al.}~\cite{Peng15} studied the trend of the free energy, by simulating BCP thin films on saw-toothed substrates. They obtained that the free energy decreases as the film thickness increases. The increase of film thickness decreases the fraction of non-flat substrate, and is equivalent to a decrease of the substrate roughness.

In conclusion, within the framework of SCFT we explored the influence of substrate roughness on the relative stability of cylindrical BCP phases having different orientations. When the substrate roughness and surface preference are too large, they will induce defects in the cylindrical phase. The impact of the amplitude, periodicity as well as the surface preference on the morphologies is detailed in our paper. Increasing the substrate roughness (increasing $R$ or separately, decreasing $L_s$) causes the cylindrical phase to prefer to have an orientation perpendicular to the substrate when the substrate preference parameter is moderate. Phase transitions from C$_{\perp}$ to C$^{2}_{\parallel}$ and, from C$^{2}_{\parallel}$ to C$^{1}_{\parallel}$ are observed via increasing the substrate preference. We find that the perpendicular cylindrical phase is more stable when the substrate corrugations undulate along the larger $b$-direction rather than along the shorter $a$-direction. Our results are seemingly robust as they in agreement with several experimental results. In addition to the roughness and preference of substrate, there are several other parameters that can influence the orientation and relative stability of the cylindrical phase. For example, the film thickness, the relative preference of the top surface, etc. We hope that our results can become a useful guide for future experiments, as well as for applications.

\bigskip
{\bf Acknowledgement.}~~
We would like to thank A.-C. Shi for useful discussions. This work was supported in
part by grants No.~21822302 and 21434001 of the National Natural Science Foundation of China (NSFC),
and the NSFC-ISF Research Program, jointly funded by the NSFC under grant No.~51561145002, and the Israel Science Foundation (ISF) under grant No.~885/15.

\newpage

%

\end{document}